\documentclass{moriond}
\usepackage{amsmath,amssymb,latexsym,amsthm,amsfonts,dsfont,cancel,braket,parskip}
\usepackage{tikz} 
\usepackage{tikz-feynman, contour}
\tikzfeynmanset{compat=1.1.0}

\usetikzlibrary{shapes,arrows,positioning,automata,backgrounds,calc,er,patterns}
\def\prd{\ref@jnl{Phys.~Rev.~D}}        % Physical Review D

\newcommand{\td}[1]{
	\if\notesOn1
	\todo[inline]{#1}
	\fi
}

\tikzset{
	graviton/.style={
		double,
		decoration={snake, aspect=0.75, mirror, segment length=1.5mm},
		decorate
	}
}

\tikzfeynmanset{
	bigblob/.style={
		shape=circle,
		draw=blue,
		fill=red}
}

\newcommand{\Photo}{}
\begin{document}
\title{Amplitudes, Gravity and Classical Discontinuities}
%\email{burgerj.daan@gmail.com}\email{raul.carballorubio@sissa.it}\email{nathantmoynihan@gmail.com}\email{jeff.murugan@uct.ac.za}\email{amanda.weltman@uct.ac.za} 
\author{Daniel J. Burger$^{(a,b)}$, Ra{\'u}l Carballo-Rubio$^{(c)}$,  
	Nathan Moynihan$^{(a,b)}$, Jeff Murugan$^{(a,b)}$
	and Amanda Weltman$^{(b),~}$}
\address{
	$ {}^{(a)}$The Laboratory for Quantum Gravity \& Strings\\
	$ {}^{(b)}$Department of Mathematics \& Applied Mathematics,
University Of Cape Town, Private Bag, Rondebosch, 7701, South Africa\\	
	$ {}^{(c)}$SISSA, International School for Advanced Studies,\\ Via Bonomea 265, 34136 Trieste, Italy
}

\maketitle\abstracts{
	On-shell methods have revitalized interest in scattering amplitudes which have, in turn, shed some much needed light on the structure of quantum field theories. These developments have been warmly embraced by the particle physics community. Less so in the astrophyical and cosmological contexts. As part of an effort to address this imbalance, we illustrate these methods by revisiting two classic problems in gravity: gravitational light-bending and the vDVZ discontinuity of massive gravity.}
	%Recursion is pointless\cite{pointless}.

	\section{Introduction} 
	The detection of gravitational waves by LIGO in 2017 has stimulated renewed interest in the calculation of observables in gravitational theories. However, in perturbative gravity this can be a cumbersome endeavour. Treating GR as a QFT can help, since classical aspects of gravity can be extracted directly from the scattering amplitudes in the $\hbar\longrightarrow 0$ limit. However, the traditional computational device - Feynman diagrams - comes with much tedious baggage; namely that we demand gauge freedom and locality be manifest at every intermediate step. This doesn't help where gravity is concerned. For example, the (off-shell) GR three-point alone is infamously 171 terms long. In the last two decades, \textit{on-shell methods} have revolutionised the computation of scattering amplitudes. Key to this program is the idea that the S-matrix ought to be computable using only physical information gained from asymptotic states. This same principle -- and on-shell methods -- can be applied to gravitational phenomena, introducing incredible simplicity and yielding some remarkable new results.
	%\blindtext
	
	%----------------------------------------------------------------------------------------
	%	BCFW RECURSION
	%----------------------------------------------------------------------------------------
	
	\section{On-Shell Methods and BCFW Recursion} Tree-level scattering amplitudes can be almost completely determined from their analytic properties, Poincar\'e invariance, dimensional analysis and a little complex analysis. Indeed, locality dictates that these amplitudes have simple poles in $p^2$. By complexifying the external momenta we can use Cauchy's theorem to build up $n$-point amplitudes from one basic building block: the on-shell three-point amplitude. This, in turn, is fixed entirely by dimensional analysis (in 4D, three-points are required to have mass dimension 1) and the knowledge of how amplitudes should transform under the little group. BCFW recursion hinges on the fact that tree amplitudes are \textit{singular} on-shell and behave as $\sim 1/\hat{P}^2$. Shifting the external momenta into the complex plane $P_i\longrightarrow P_i + z\eta_i$ produces a simple pole in $z$. The amplitude, $A(z)$ is now rational functions of $z$ which submits to Cauchy's theorem. Knowing only the \textit{poles} and \textit{residues} we can write,
	$\displaystyle
	A_n = i\sum_{z_{ib}}\sum_{h}A_L(z_{ib})\frac{1}{P_{ib}^2}A_R(z_{ib}).\label{eq:recrel}
	$
	Amazingly, we can build \textit{any} tree level amplitude from \textit{on-shell} 3-point + poles! In pictures,
	\begin{equation*}
	\begin{gathered}
	\begin{tikzpicture}[scale=0.5]
	\begin{feynman} 
	\vertex (a) at (-2,1);
	\vertex (b) at (-2,-1);
	\vertex (c) at (2,-1);
	\vertex (d) at (2,1);
	\vertex (p) at (0,0);
	\vertex[dot] (ap) at (-2.1,1.3);
	\diagram* {
		(a) -- [plain] (p) -- [plain] (d),
		(b) -- [plain] (p) -- [plain] (c)
	};
	\draw[preaction={fill, white},pattern=north east lines] (0,0) ellipse (0.8cm and 0.8cm);
	\filldraw (-1.8,0.35) circle (1.5pt);
	\filldraw (-1.8,-0.35) circle (1.5pt);
	\filldraw (1.8,0.35) circle (1.5pt);
	\filldraw (1.8,-0.35) circle (1.5pt);
	\filldraw (2,0) circle (1.5pt);
	\filldraw (-2,0) circle (1.5pt);
	%\draw[preaction={fill, white},pattern=north east lines] (1.9,-0.25) ellipse (0.3cm and 0.8cm);
	%\draw [densely dashed, red, line width=0.5mm,] (0,1.2) -- (0,-1.6);
	%\draw (-1,0.1) node[above] {$\ell_1 \longrightarrow$};
	%\draw (-1,-0.65) node[below] {$\ell_2 \longrightarrow$};
	\end{feynman}
	\end{tikzpicture}
	\end{gathered}
	~~~=~~~
	\begin{gathered}
	\begin{tikzpicture}[scale=0.5]
	\begin{feynman} 
	\vertex (a) at (-2,1);
	\vertex (b) at (-2,-1);
	\vertex (c) at (2,-1);
	\vertex (d) at (2,1);
	\vertex (p) at (-1,0);
	\vertex (q) at (1,0);
	\diagram* {
		(a) -- [plain] (p) -- [plain, edge label'=\(1/z\)] (q) -- [plain] (d),
		
		(b) -- [plain] (p) -- [plain] (q) -- [plain] (c)
	};
	\draw[preaction={fill, white},pattern=north east lines] (-1,0) ellipse (0.5cm and 0.5cm);
	\draw[preaction={fill, white},pattern=north east lines] (1,0) ellipse (0.5cm and 0.5cm);
	\filldraw (-1.8,0.35) circle (1.5pt);
	\filldraw (-1.8,-0.35) circle (1.5pt);
	\filldraw (1.8,0.35) circle (1.5pt);
	\filldraw (1.8,-0.35) circle (1.5pt);
	\filldraw (2,0) circle (1.5pt);
	\filldraw (-2,0) circle (1.5pt);
	\end{feynman}
	\end{tikzpicture}
	\end{gathered}
	\end{equation*}
	We can use symmetry to fix three-points uniquely, using our knowledge of the \textit{little group}: $p_i^\mu \longrightarrow L^\mu_{~\nu} p_i^\nu = p_i^\mu$. Using little group adapted variables (i.e. noting that $SO(1,3)\simeq SL(2,\mathbb{C})$), we note that amplitudes transform under $U(1)$ and $SU(2)$ respectively. Together with dimensional analysis, these results allow for the unique (up to a coupling constant) construction of on-shell three-point amplitudes for any spin. Remarkably, this means that $n$-point tree-level amplitudes involving massless particles of spin $s \leq 2$ can be built entirely without the baggage of a Lagrangian, field redefinitions or gauge symmetry (modulo some technicalities).
	
	%----------------------------------------------------------------------------------------
	%	MATERIALS AND METHODS
	%----------------------------------------------------------------------------------------
	
	%\section{Gravitational Light-bending}
	\section{Gravitational Light-bending} A key prediction of GR is the classical deflection of light due to a massive body, like a star or black hole. As a scattering problem, we can think of this simply as the deflection of a massless particle by a massive one, mediated by a graviton. %In terms of Feynman diagrams:
	
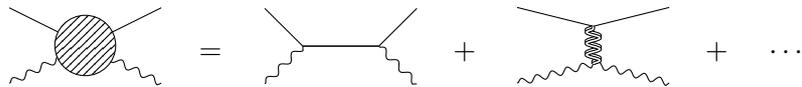
\begin{equation*}
	\begin{gathered}
	\begin{tikzpicture}[scale=0.5]
	\begin{feynman} 
	\vertex (a) at (-2,1);
	\vertex (b) at (-2,-1);
	\vertex (c) at (2,-1);
	\vertex (d) at (2,1);
	\vertex (p) at (0,0);
	\vertex[dot] (ap) at (-2.1,1.3);
	\diagram* {
		(a) -- [plain] (p) -- [plain] (d),
		(b) -- [photon] (p) -- [photon] (c)
	};
	\draw[preaction={fill, white},pattern=north east lines] (0,0) ellipse (0.8cm and 0.8cm);
	%\draw[preaction={fill, white},pattern=north east lines] (1.9,-0.25) ellipse (0.3cm and 0.8cm);
	%\draw [densely dashed, red, line width=0.5mm,] (0,1.2) -- (0,-1.6);
	%\draw (-1,0.1) node[above] {$\ell_1 \longrightarrow$};
	%\draw (-1,-0.65) node[below] {$\ell_2 \longrightarrow$};
	\end{feynman}
	\end{tikzpicture}
	\end{gathered}
	~~~=~~~
	\begin{gathered}
	\begin{tikzpicture}[scale=0.5]
	\begin{feynman} 
	\vertex (a) at (-2,1);
	\vertex (b) at (-2,-1);
	\vertex (c) at (2,-1);
	\vertex (d) at (2,1);
	\vertex (p) at (-1,0);
	\vertex (q) at (1,0);
	\diagram* {
		(a) -- [plain] (p) -- [plain] (q) -- [plain] (d),
		
		(b) -- [photon] (p) -- [plain] (q) -- [photon] (c)
	};
	\end{feynman}
	\end{tikzpicture}
	\end{gathered}
	~~~+~~~
	\begin{gathered}
	\begin{tikzpicture}[scale=0.5]
	\begin{feynman} 
	\vertex (a) at (-2,1);
	\vertex (b) at (-2,-1);
	\vertex (c) at (2,-1);
	\vertex (d) at (2,1);
	\vertex (p) at (0,0.5);
	\vertex (q) at (0,-0.5);
	\diagram* {
		(a) -- [plain] (p) -- [graviton] (q) -- [photon] (c),
		
		(b) -- [photon] (q) -- [graviton] (p) -- [plain] (d),
	};
	\end{feynman}
	\end{tikzpicture}
	\end{gathered}
	~~~+~~~
	\begin{gathered}
	%\begin{tikzpicture}[scale=1.8]
	%\begin{feynman} 
	%\vertex (a) at (-2,1);
	%\vertex (b) at (-2,-1);
	%\vertex (c) at (2,-1);
	%\vertex (d) at (2,1);
	%\vertex (p) at (0,0);
	%\vertex (q) at (0,0);
	%\diagram* {
	%	(a) -- [plain] (p) -- [graviton] (q) -- [photon] (c),
	%	
	%	(b) -- [photon] (q) -- [graviton] (p) -- [plain] (d),
	%};
	%\end{feynman}
	%\end{tikzpicture}
	\cdots
	\end{gathered}
	\end{equation*}
	
	Computing these via the usual Feynman diagram method is possible, but using the BCFW recursion relations one can choose the complex shifts in such a way that the all the information of the interaction is contained in the diagram with the scalar propagator \cite{Burger:2017yod}. This makes the computation of the amplitude trivial as we only need the on-shell three points $M_3[1^0,2^{-1},P^{0}]$, $M_3[3^0,4^{-1},P^{0}]$ and the scalar propagator. Constructing the relevant three-points using symmetry arguments and BCFW, we find that the amplitude is given by
        \begin{eqnarray*}
	A_4[1^{+1},2^{-1},3,4] &=& A[1^+,\hat{2}^-,\hat{P}^{-2}_{12}]\frac{1}{P_{12}^2}A[-\hat{P}^{+2}_{12},\hat{3},4] = \frac{\kappa^2}{4}\frac{\bra{2}p_4|1]^2}{P_{12}^2}.\label{eq:nn4}
	\end{eqnarray*}
	To compare this with the classical result, we need to take the appropriate limits of the amplitude, specifically the limit that gives the small angle approximation, in this case $m_\phi \gg E_\gamma$ and small momentum transfer $t$. Computing the cross-section then leads directly to the scattering angle predicted by GR. One might ask what would happen if we were to replace the photon by a graviton, probing how a gravitational wave might be deflected by a massive body. The equivalence principle demands that the bending angle ought to be the same in both cases, and yet the amplitudes are very different off-shell, differing by 100's of terms due to the complicated graviton three-point function. Thankfully, it is barely any more complicated when using on-shell methods and BCFW and we find $A[1^{+2},2^{-2},3,4]=\frac{i\kappa^2}{16}\frac{\bra{1}p_4|2]^4}{P_{12}^2}\left(\frac{1}{(p_2\cdot p_3)(p_2\cdot p_4)}\right)$.
	This is obviously a different amplitude, but it too leads to the correct bending angle. This is because  of the observations that $|A[1^{+2},2^{-2},3,4]|^2 = |A[1^{+1},2^{-1},3,4]|^2f(s_{12},s_{13},s_{14})^2$ and $f(s_{12},s_{13},s_{14})\bigg|_{t < 1,m_\phi \gg E_\gamma} \simeq 1$. 
	
	% which is (1) the mass of the scalar is much larger than the energy of the photon and (2) there is small momentum transfer between the photon and scalar. It is also good to note that we do not consider multiple graviton exchange in our computation.
	
	\section{The vDVZ Discontinuity Revisited} The vDVZ discontinuity manifests itself as an intriguing puzzle: classical gravity with a non-zero massive graviton does not yield the results of GR in the massless limit. Typically, this is seen in the field theory itself, and can be made apparent either at the level of the Lagrangian or by computing the propagator. Observationally, this results in the light-bending angle being rescaled by a factor of $3/4$ -- an extremely unwelcome addition. Can we see this purely \textit{on-shell}? Indeed we can \cite{Moynihan:2017tva}. Compare the following amplitudes
	
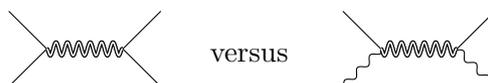
\begin{equation*}\label{key}
	\begin{gathered}
	\begin{tikzpicture}[scale=0.5]
	\begin{feynman} 
	\vertex (a) at (-2,1);
	\vertex (b) at (-2,-1);
	\vertex (c) at (2,-1);
	\vertex (d) at (2,1);
	\vertex (p) at (-1,0);
	\vertex (q) at (1,0);
	\diagram* {
		(a) -- [plain] (p) -- [graviton] (q) -- [plain] (d),
		
		(b) -- [plain] (p) -- [graviton] (q) -- [plain] (c)
	};
	\end{feynman}
	\end{tikzpicture}
	\end{gathered}
	~~~~~\text{versus}~~~~~
	\begin{gathered}
	\begin{tikzpicture}[scale=0.5]
	\begin{feynman} 
	\vertex (a) at (-2,1);
	\vertex (b) at (-2,-1);
	\vertex (c) at (2,-1);
	\vertex (d) at (2,1);
	\vertex (p) at (-1,0);
	\vertex (q) at (1,0);
	\diagram* {
		(a) -- [plain] (p) -- [graviton] (q) -- [plain] (d),
		
		(b) -- [photon] (p) -- [graviton] (q) -- [photon] (c)
	};
	\end{feynman}
	\end{tikzpicture}
	\end{gathered}
	\end{equation*}
	Since now we are dealing with amplitudes of massive particles, they need to have indices that correspond to their $SU(2)$ transformation properties. Hence, $ M_{4,t}^{all~scalar} = \frac{M_3^{IJKL}\overline{M}_{3,IJKL}}{t}$, with $M_3^{IJKL}$ entirely fixed by $SU(2)$ and dimensional analysis.  Taking the Newtonian limit of $M_{4,t}^{all~scalar}$ we find the potential in momentum space $T_{fi}^{COM}(0) = \frac{4}{3}\left(\frac{4\pi Gm_\phi^2}{\vec{q}^{\,\,2}}\right)\,.$ This is however not the correct Newtonian limit, so we rescale $G\longrightarrow\tilde{G} = \frac{3}{4}G$ and all seems well. Computing the photon case yields
       \begin{eqnarray*}
          M_{4,t}^{photon-scalar} = \frac{M_3^{IJKL}\overline{M}_{3,IJKL}}{t} = M_{4,t}^{photon-scalar}\bigg|_{m_g = 0}\,.
       \end{eqnarray*}   
	Including the massive graviton changes nothing in the case of photons! However, deriving the light bending angle as before, we find that $\theta_{m_g\longrightarrow 0} = \frac{3}{4}\theta_{m_g = 0}$, manifesting the vDVZ discontinuity.
	
	%----------------------------------------------------------------------------------------
	%	FORTHCOMING RESEARCH
	%----------------------------------------------------------------------------------------
	
	\section{A Spin 3/2 Discontinuity} It is natural to wonder whether this discontinuity persists under a supersymmetry transformaion: does a similar discontinuity exist for the massive gravitino in $\mathcal{N} = 1$ SUGRA? We can again consider comparing two diagrams, now considering the interaction of light with fermionic matter. The two amplitudes we must compare are
	
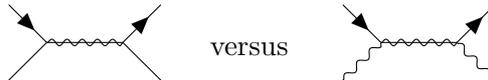
\begin{equation*}\label{key}
	\begin{gathered}
	\begin{tikzpicture}[scale=0.5]
	\begin{feynman} 
	\vertex (a) at (-2,1);
	\vertex (b) at (-2,-1);
	\vertex (c) at (2,-1);
	\vertex (d) at (2,1);
	\vertex (p) at (-1,0);
	\vertex (q) at (1,0);
	\diagram* {
		(a) -- [fermion] (p) -- [photon] (q) -- [fermion] (d),
		
		(b) -- [plain] (p) -- [plain] (q) -- [plain] (c)
	};
	\end{feynman}
	\end{tikzpicture}
	\end{gathered}
	~~~~~\text{versus}~~~~~
	\begin{gathered}
	\begin{tikzpicture}[scale=0.5]
	\begin{feynman} 
	\vertex (a) at (-2,1);
	\vertex (b) at (-2,-1);
	\vertex (c) at (2,-1);
	\vertex (d) at (2,1);
	\vertex (p) at (-1,0);
	\vertex (q) at (1,0);
	\diagram* {
		(a) -- [fermion] (p) -- [photon] (q) -- [fermion] (d),
		
		(b) -- [photon] (p) -- [plain] (q) -- [photon] (c)
	};
	\end{feynman}
	\end{tikzpicture}
	\end{gathered}
	\end{equation*}
	%We are making a study of seeing how this discontinuity manifests in the massive amplitude formalism and nearing completion. To this end we need to compute the following amplitudes,
	%
	%\begin{align*}
	%	M_4({\bf 1}^{\pm\frac12},2^{-1},{\bf 3}^{\mp\frac12},4^{+1}) = 
	%	\begin{gathered}
	%	\begin{tikzpicture}[scale=1.8]
	%	\begin{feynman} 
	%	\vertex (a) at (-2,1);
	%	\vertex (b) at (-2,-1);
	%	\vertex (c) at (2,-1);
	%	\vertex (d) at (2,1);
	%	\vertex (p) at (-1,0);
	%	\vertex (q) at (1,0);
	%	\diagram* {
	%		(a) -- [plain] (p) -- [plain] (q) -- [plain] (d),
	%		(p) -- [photon] (q),
	%		(b) -- [photon] (p) -- [plain] (q) -- [photon] (c)
	%	};
	%	\end{feynman}
	%	\end{tikzpicture}
	%	\end{gathered} \\
	%	M_4({\bf 1}^{\pm\frac12},2^{0},{\bf 3}^{\mp\frac12},4^{0}) = 
	%	\begin{gathered}
	%	\begin{tikzpicture}[scale=1.8]
	%	\begin{feynman} 
	%	\vertex (a) at (-2,1);
	%	\vertex (b) at (-2,-1);
	%	\vertex (c) at (2,-1);
	%	\vertex (d) at (2,1);
	%	\vertex (p) at (-1,0);
	%	\vertex (q) at (1,0);
	%	\diagram* {
	%		(a) -- [plain] (p) -- [plain] (q) -- [plain] (d),
	%		(p) -- [photon] (q),
	%		(b) -- [dashed] (p) -- [plain] (q) -- [dashed] (c)
	%	};
	%	\end{feynman}
	%	\end{tikzpicture}
	%	\end{gathered}.
	%\end{align*}

	In both cases we keep the external fermions massive, but the scalars and photons massless. In the graviton case, we found that the discontinuity arose due to the non-vanishing contribution from the dilaton. In the case of the gravitino, comparing the amplitudes above, we find a similar result: the amplitudes involving scalars not equal in the massless limit of the Rarita-Schwinger field, while the amplitude involving photons is equal. This time, the discontinuity arises from the goldstino-mode of the Rarita-Schwinger field, i.e. the spin-$1/2$ mode of the gravitino. We will explore this result in more detail in a forthcoming paper. 
	
\textbf{\textit{Acknowledgements:}} AW \& NM are supported by the South African Research Chairs Initiative of the Department of Science and Technology and the National Research Foundation of South Africa. DJB is supported by a PHD fellowship from the South African National Institute for Theoretical Physics. JM is supported by the NRF of South Africa under grant CSUR 114599.


\section*{References}
\begin{thebibliography}{99}
%\cite{Burger:2017yod}
\bibitem{Burger:2017yod} 
  D.~J.~Burger, R.~Carballo-Rubio, N.~Moynihan, J.~Murugan and A.~Weltman,
  ``Amplitudes for astrophysicists: known knowns,''
  Gen.\ Rel.\ Grav.\  {\bf 50}, no. 12, 156 (2018)
  doi:10.1007/s10714-018-2475-0
  [arXiv:1704.05067 [astro-ph.HE]].
  %%CITATION = doi:10.1007/s10714-018-2475-0;%%
  %5 citations counted in INSPIRE as of 12 May 2019
%\cite{Moynihan:2017tva}
\bibitem{Moynihan:2017tva} 
  N.~Moynihan and J.~Murugan,
  ``Comments on scattering in massive gravity, vDVZ and BCFW,''
  Class.\ Quant.\ Grav.\  {\bf 35}, no. 15, 155005 (2018)
  doi:10.1088/1361-6382/aacb73
  [arXiv:1711.03956 [hep-th]].
  %%CITATION = doi:10.1088/1361-6382/aacb73;%%
  %1 citations counted in INSPIRE as of 12 May 2019  
\end{thebibliography}
\end{document}